\documentclass{article}

\usepackage{PRIMEarxiv}

\usepackage[utf8]{inputenc} 
\usepackage[T2A]{fontenc}    
\usepackage[russian,english]{babel}
\usepackage{hyperref}
\usepackage{natbib}
\usepackage{url}            
\usepackage{booktabs}       
\usepackage{amsfonts}       
\usepackage{nicefrac}       
\usepackage{microtype}      
\usepackage{lipsum}
\usepackage{fancyhdr}       
\usepackage{graphicx}       
\usepackage{makecell}
\usepackage{caption}
\usepackage{longtable}
\graphicspath{{media/}}     

\title{\textbf{Mapping Literary Space: \\A Social Network from the Timeline of Cultural Events}
}

\author{
  Maria Levchenko \\
  University of Bologna \\
  \texttt{maria.levchenko@studio.unibo.it} \\
}

\begin{document}
\maketitle

\begin{abstract}
This study applies social network analysis (SNA) to map and analyze literary networks in St Petersburg from 1999 to 2019, using data from the 'SPbLitGuide' newsletter. By examining co-participation in literary events, we reveal the dynamics and structures of these networks, identifying key communities and influential figures. Our network graph, consisting of 14,066 nodes and 127,068 edges, represents a highly interconnected and cohesive small-world network with robust local clustering and extensive collaboration. Focusing on core participants, we refined the graph and applied community detection methods to identify distinct groups with specific aesthetic preferences and personal connections. 
\end{abstract}


Exploring the dynamics of literary networks and identifying communities based on a chronology of cultural events and their textual representations is a novel approach in the field of digital humanities. While the chronology of cultural events has been a focus for scholars of Russian literature, the study of event announcements in sociological studies of contemporary literature is relatively new. The application of text processing techniques and SNA in this context holds promising potential for advancing research.

Following the seminal work of \citet{Moretti2005} and \citet{Elson2010}, SNA has become a common tool for analyzing literary networks. Studies such as \citet{So2013} have objectively revealed complex literary community networks, while recent work by \citet{RoigSanz2021} has explored literary cross-referencing and translation. Building on these studies, our research applies SNA to event data on contemporary Russian culture, extracting actors' connections through event participation.

This study uses a dataset parsed from the "SPbLitGuide" (Saint Petersburg Literary Guide), an ongoing newsletter launched by Daria Sukhovey in May 1999 that chronicles literary events in Saint Petersburg. The SPB LitGuide provides detailed information about upcoming events, including dates, times, locations, brief descriptions, and details about the organizers and key participants, such as hosts and speakers. 

\begin{figure}
    \centering
    \includegraphics[width=0.9\linewidth]{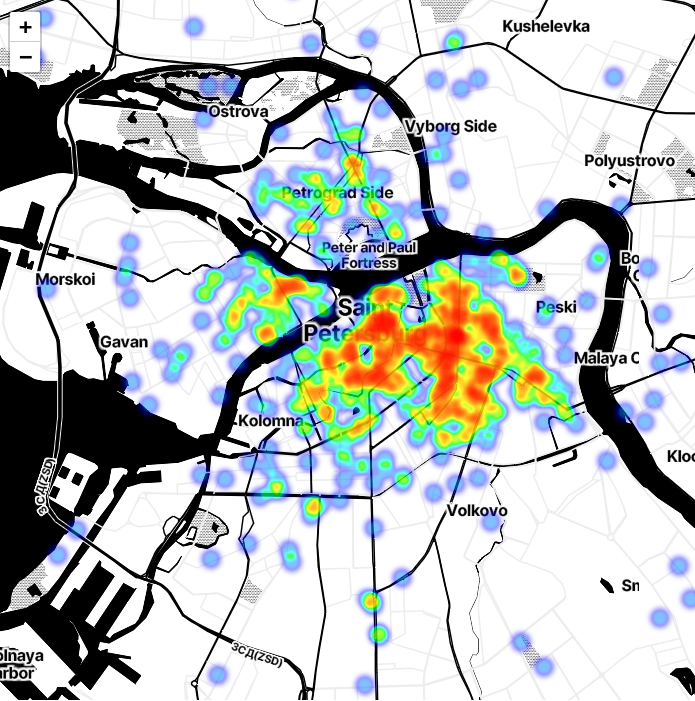}
    \caption{Event Frequency Heat Map}
    \label{fig:heatmap}
\end{figure}

Our dataset includes all issues of SPbLitGuide from May 1999 to October 2019, covering 1,255 issues and approximately 14,990 events in 862 locations. The data was parsed into CSV format, structured using Natasha and DeepPavlov, and enriched with geographic data using the Yandex Geocoder API. The dataset, available on Zenodo \citet{Levchenko2023}, allowed us to generate heatmaps and network graphs of literary events in St Petersburg (see Figure \ref{fig:heatmap}).

Announcements for literary events are pragmatic and aimed at potential attendees. They tend to be short and structured, giving the date, time, venue and a brief description to inform and attract visitors. The authors of these descriptions are usually anonymous, but are likely to be the event organisers or venue owners. While the length of the descriptions can vary, they generally provide enough information for readers to understand the event and its participants.

The event descriptions contain three distinct categories of names:
1. The names of actual participants in the events.
2. The names of individuals to whom the events are dedicated.
3. The names of individuals connected to the actual participants, such as colleagues, friends, or critics. These names are included for the purpose of providing information about the main figures involved in the events.

This research focuses on the active participants mentioned in the descriptions, such as organisers, speakers and presenters. It is obvious that co-participation in the event suggests connections between individuals, implying that they may belong to the same literary groups, be involved in the same cultural project, or have similar aesthetic preferences. Thus, the analysis of co-participation data can reveal the social structure and dynamics within the literary field.

\begin{figure*}[h!]
\centering
\begin{minipage}{.45\textwidth}
  \centering
  \includegraphics[width=\linewidth]{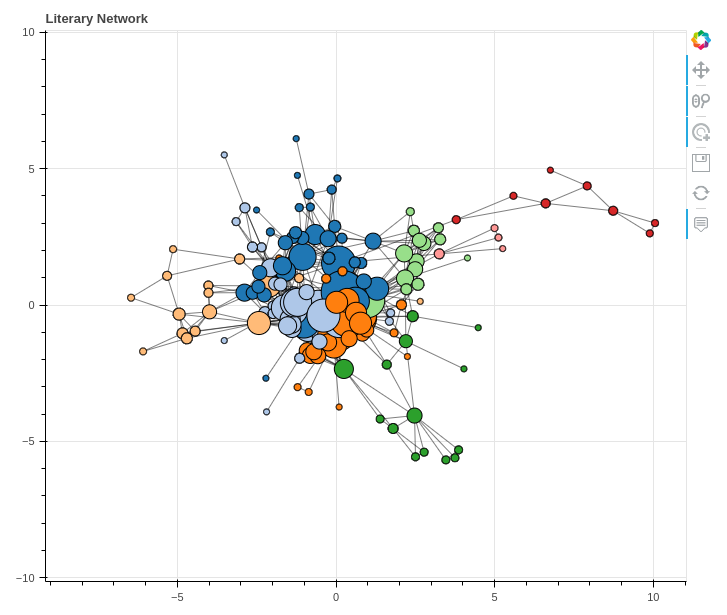}
  \captionof{figure}{The literary network for the year 2019}
  \label{fig:network}
\end{minipage}%
\hfill
\begin{minipage}{.45\textwidth}
  \centering
  \label{tab:table1}
  \begin{tabular}{ll}
    \hline
    \textbf{Parameter} & \textbf{Value} \\
    \hline
    Number of Nodes & 14,066 \\
    Number of Edges & 127,068 \\
    Connected Components & 466 \\
    Average Clustering & 0.751 \\
    Diameter & 13 \\
    Density & 0.0013 \\
    Coverage & 0.808 \\
    Performance & 0.875 \\
    Average Path Length & 3.865 \\
    Global Efficiency & 0.227 \\
    \hline
  \end{tabular}
  \captionof{table}{Key parameters of the entire network}
  \label{tab:table_1}
\end{minipage}
\end{figure*}

To map the connections between literary actors, we used the co-participation data to create a network graph. This graph shows the relationships and collaborations between literary actors in St. Petersburg, year by year (see Figure \ref{fig:network}). Each point or "node" represents a literary actor, while the connecting lines or "edges" represent their collaborative participation in events.

The initial graph, based on event co-participation, contained 14,066 nodes and 127,068 edges across 466 components, revealing a highly interconnected structure. Key parameters such as clustering and density were analyzed (see Table \ref{tab:table_1}), indicating a cohesive structure with robust local clustering and extensive collaboration, characteristic of a resilient small-world network. To focus on core actors (those involved in five or more events), we filtered the dataset, resulting in a refined graph with 1,609 nodes and 35,803 edges. We used greedy modularity maximization and the Louvain method for community detection. The main parameters of the detected communities are summarised in Table \ref{table:community_parameters}.

This approach maps the existing literary networks and lays the groundwork for future research and analysis, providing insights into the evolving nature of literary collaborations and communities in St Petersburg. Future research could verify these findings by comparing them with connections extracted from Wikipedia or social media links, as key figures are often represented on these platforms. In addition, extending the analysis to other levels of connections, such as name-dropping and mentions of non-participants, could shed light on other trends in the dynamics of the literary field.

\begin{longtable}{|c|r|r|r|r|p{6cm}|}
\caption{Key Parameters of Detected Communities} \label{table:community_parameters} \\
\hline
\textbf{Community} & \textbf{Size} & \textbf{Diameter} & \makecell{\textbf{Average} \\ \textbf{Clustering}} & \textbf{Density} & \textbf{Top Nodes by Degree Centrality} \\
\hline
\endfirsthead
\caption[]{Key Parameters of Detected Communities (continued)} \\
\hline
\textbf{Community} & \textbf{Size} & \textbf{Diameter} & \makecell{\textbf{Average} \\ \textbf{Clustering}} & \textbf{Density} & \textbf{Top Nodes by Degree Centrality} \\
\hline
\endhead
\hline \multicolumn{6}{|r|}{{Continued on next page}} \\ \hline
\endfoot
\hline
\endlastfoot
C3 «Maitres»       & 563           & 10                & 0.99                       & 0.0275           & Pavel Krusanov (289), Sergey Stratanovsky (277), Sergey Nosov (272), Valery Popov (252), Valery Shubinsky (224), Yakov Gordin (220), Aleksandr Sekatsky (216), Aleksandr Kushner (208), Andrey Aryev (198), Irina Dudina (176) \\ \hline
\multicolumn{6}{|p{16cm}|}{\textbf{The largest community with moderate clustering and low density, indicating a dispersed and loosely connected structure. The most influential members are particularly well-known figures from the 1960s, embodying the convergence of the official and underground cultural spheres during the late Soviet era.}} \\ \hline
C4 «Performancers» & 408           & 1                 & 1.00                       & 1.0000           & Vladimir Antipenko (437), Maria Agapova (325), Mikhail Solonnikov (306), Tibul Kamchatsky (240), Ilya Zhigunov (228), Maria Amfilokhieva (208), Alla Zinevich (196), Igor Nikolsky (194), Dmitry Ademin (194), Svyatoslav Korovin (189) \\ \hline
\multicolumn{6}{|p{16cm}|}{\textbf{This community stands out as a fully connected clique, which could indicate a highly cohesive subgroup or a tightly knit cluster within the network. The main figures of this community are grouped around the poet theater "Hear!"}} \\ \hline
C0 «Intellectuals» & 362           & 5                 & 0.0087                     & 0.0607           & Arsen Mirzaev (1300), Dmitry Grigoryev (1225), Valery Zemskikh (1185), Darya Sukhovey (1165), Aleksandr Skidan (868), Tamara Bukovskaya (801), Dmitry Chernyshov (712), Valery Mishin (706), Nikita Safonov (640), Nastya Denisova (621) \\ \hline
\multicolumn{6}{|p{16cm}|}{\textbf{A moderately sized community with a compact structure but very low clustering, suggests a more distributed and less cohesive structure. These members, who began their careers in the late 1990s, now play an important role in St. Petersburg's cultural sphere, especially in avant-garde expressions. They are indicative of well-educated poets and philosophers who not only contribute to the community's dynamism but also reflect its intellectual and innovative character.}} \\ \hline
C2 «Traditionalists» & 260         & 5                 & 0.0343                     & 0.1013           & Galina Ilyukhina (472), Dmitry Legeza (459), Evgeny Myakishev (456), Nina Savushkina (285), Aleksandr Frolov (270), Aleksandr Dzhigit (259), Evgeny Antipov (251), Olga Khokhlova (243), Vyacheslav Leykin (241), Vsevolod Gurevich (224) \\ \hline
\multicolumn{6}{|p{16cm}|}{\textbf{The smallest community with moderate density and low clustering, indicates some internal connectivity but is still relatively sparse. The members of this community are particularly active in traditional poetry circles, grouped around the literary group "LITO Piiter".}} \\ \hline
\end{longtable}

\end{document}